\documentclass[intlimits,twoside,a4paper]{article}

\usepackage{cite}
\usepackage{multicol}
\usepackage{multirow}
\usepackage{subfig}
\usepackage[cp1251]{inputenc}

\usepackage[eqsecnum]{cmpj3}

\usepackage{bm}

\issue{2023}{26}{2}{23701}
\doinumber{10.5488/CMP.26.23701}

\title[ Computational study of structural, elastic, electronic]%
{Computational study of structural, elastic, electronic, phonon dispersion relation and thermodynamic properties of orthorhombic 
CaZrS$_3$ for optoelectronic applications}%

\author[M. D. Kassa,  N. G. Debelo, M. M. Woldemariam] {{M. D. Kassa\orcid{0000-0002-5524-6791}}, 
    {N. G. Debelo,}
    {M. M. Woldemariam\thanks{Corresponding author: Menberu Mengesha Woldemariam, 
    		email: \email{menberu.mengesha@ju.edu.et}.}}}
\address{
 Department of Physics,   Jimma University, P. O. Box 378, Jimma, Ethiopia.\\
  \email{mulugettaduressa2020@gmail.com}, \email{nafsif@gmail.com}
}

\Keywords{CaZrS$_3$, electronic property, mechanical property, phonon dispersion, thermodynamic property}

\date{Received August 02, 2022, in final form September 26, 2022}

\begin{document}

\maketitle

\begin{abstract}
Chalcogenide perovskites offer superior thermal and aqueous stability as well as a benign elemental composition compared to organic halide perovskites for optoelectronic applications. In this study, the structural, electrical, elastic, phonon dispersion, and thermodynamic features of the orthorhombic phase of chalcogenide perovskite CaZrS$_3$ (space group \emph{Pnma}) were examined by first principles calculations utilizing the plane wave pseudopotentials~(PW-PPs)  in generalized gradient approximations~(GGA). The ground state properties such as lattice parameters, unit cell volume, bulk modulus, and its derivative were calculated and are in a good agreement with existing findings. The mechanical properties such as bulk modulus, shear modulus, Young's modulus and elastic anisotropy were calculated from the obtained elastic constants. The ratio of bulk modulus to shear modulus confirms that the orthorhombic phase of CaZrS$_3$ is a ductile material. The absence of negative frequencies in phonon dispersion curve and the phonon density of states give an indication that the structure is dynamically stable. Finally, thermodynamic parameters such as free energy, entropy, and heat capacity were calculated with variation in temperature. The estimated findings follow the same pattern as previous efforts.
%
%
\printkeywords
%
\end{abstract}

\section{Introduction}


The quest for technological advancement, particularly in the field of semiconductors, plays a significant role in several optoelectronic, photonic, and energy technologies. Among current semiconductors, the prevailing materials like silicon, group III--V, and group II--VI are typically constructed by a fourfold coordinated tetrahedral network of covalent bonds. There have been major successes in developing solar cell semiconductor materials such as Si, GaAs, CuIn$_x$Ga$_{1-x}$Se$_2$ (CIGS), and lead halide perovskite based materials~\cite{Gre2012,Yos2017,Sna2018}. Over the past decades, inorganic-organic/organometal lead halides have been extensively studied since the early 20th century~\cite{Web1978,K2009}. In 1970, for the first time, Weber reported the synthesis and physical properties of $ -$CH$_3$NH$_3$PbX$_3-$~(X $=$ Cl, Br, I) organometal lead halide perovskite~\cite{Web1978}. During this time, the organometal lead halide perovskites emerged as a candidate and very promising materials for light harvesting in the solar cell as reported in 2009~\cite{K2009}. The latest area in solar cell materials is the organic-inorganic hybrid lead (Pb) halide perovskites, for which the efficiency reached 22.7$\%$ in~2018~\cite{Sna2018}, starting from 3.8$\%$ in 2009~\cite{K2009}.These hybrid perovskites, however, have poor thermal and moisture stability, as well as the presence of lead toxicity~\cite{Bab2016}. The intrinsic poor long-term stability of CH$_3$NH$_3$PbI$_3$-based perovskite solar cells, as well as the presence of toxicity, has hindered their industrial application and commercialization. Inorganic lead-free perovskites have recently been widely investigated to address these issues. Recently, chalcogenide perovskites have received attention due to their promising photovoltaic and thermoelectric properties, with initial studies conducted on oxide perovskites that have good band gaps for optical absorption~\cite{Gri2013}. Chalcogenide perovskites assume an ABX$_3$ configuration with~A, and B are elements with a combined valence of 6 (with different valences), while X is typically~S or~Se.  These materials belong to a new class of ionic semiconductors. The band gap of these materials can be systematically tuned in a wide range from ultra-violet to infrared. Due to their predicted strong iconicity, they may exhibit unique physical properties such as being free of deep-level defects, which is beneficial for energy harvesting and other optoelectronic applications~\cite{Per2016}. It should be emphasized that oxide perovskites, with a chemical formula ABO$_3$, have long been a focus of active research. This family of materials exhibits unusually rich properties ranging from colossal magnetoresistance, ferroelectricity to superconductivity and charge density waves, resulting from the interplay of different degrees of freedom with similar energy scales. The intriguing physics of their chalcogenide counterparts, however, is largely unexplored~\cite{Per2016}.

In recent years, theoretical calculations based on the density functional theory (DFT) have been used to reveal and predict the structural, mechanical, electrical, optical, and thermal properties of crystal materials. CaZrS$_3$, which is the focus of this work, is a family of chalcogenide perovskite crystal material that was considered theoretically for optoelectronic applications~\cite{Per2016,Elu2016,Oum2019, Zha2017, Maj2020}. The relaxed lattice parameters and band gap of CaZrS$_3$ were estimated using  DFT as implemented in VASP and the Perdew, Burk, and Ernzerhof (GGA-PBE) generalized gradient approximation~\cite{Kum2021}. In addition, the lattice parameter and the band gap of CaZrS$_3$ were also calculated using the FPLAPW method with DFT as implemented in WIEN2K, approximating the exchange correlation potential with PBE-GGA, Engel-Vosko (EV) method and Hubbard parameter (GGA+U)~\cite{Maj2020}. However, to our knowledge, the structural, elastic, electronic, phonon dispersion and thermodynamic properties of CaZrS$_3$ are not yet well investigated for optoelectronic application. Moreover, investigation of elastic, phonon dispersion and thermodynamic properties of CaZrS$_3$ using the first principle computational methods remains unexplored. In this paper, the structural, elastic, electronic, and phonon dispersion relation and thermal properties of CaZrS$_3$ are carefully examined. The electronic properties are calculated by considering PBE-GGA~\cite{Per1996} and also DFT with the Hubbard functional (DFT+U)~\cite{Flo2020} for exchange correlation potential using Quantum ESPRESSO package (QE). In addition, phonon dispersion and the thermodynamic properties of CaZrS$_3$ are studied using a $1 \times 1 \times 2$ (in $x$, $y$, and $z$ direction, respectively) supercell containing 40~atoms created in a PHONOPY package~\cite{Tog2015}.

\section{Computational methods}

In this study, the DFT as implemented in the QE~\cite{Gia2009} within the generalized gradient approximation (GGA) functional~\cite{Per1996} and with the Hubbard correction (DFT+U)~\cite{Flo2020} was used. The effective Hubbard parameter ($U_{\rm eff}$) was calculated iteratively for Zr-$d$ orbitals.  For this study, a cell with 20 atoms (4-Ca, 4-Zr, and 12-S) in orthorhombic phase for structural, elastic and electronic property calculations was used. The ultra-soft pseudopotentials~(US-PP) were used to treat the interaction of the electrons with the ion cores as in~\cite{Van1977}.  The corresponding  valence electrons considered for the calculations are Ca --- [Ar]$4s^2$, Zr --- [Kr]$4d^2 5s^2$, and S --- [Ne]$ 3s^2 3p^4$. Crystal structure optimization was done using a plane wave cutoff energy of 60 Ry and the Brillouin zone with a $3 \times 3 \times 3$ Monkhorst-Pack $k$-point grid~\cite{Mon2019} based on the convergence criteria energy $10^{-4}$~Ry, force $10^{-3}$~Ry/Bohr, and cell pressure~0.5~kbar. Using the optimized structure, the elastic properties were calculated by THERMO\_PW package within Quantum ESPRESSO package~\cite{Gia2009}.  In addition, the thermal properties were calculated with the help of PHONOPY package~\cite{Tog2015}. To study the phonon dispersion relation and thermodynamic properties of CaZrS$_3 $, a supercell of $1 \times 1 \times 2$ in $x$, $y$, and $z$ direction with 40~atoms was created and used for computations. 

\newpage

\section{Results and discussion}

\subsection{Crystal structure}

\begin{figure}[htb]
\centerline{\includegraphics[width=0.40\textwidth]{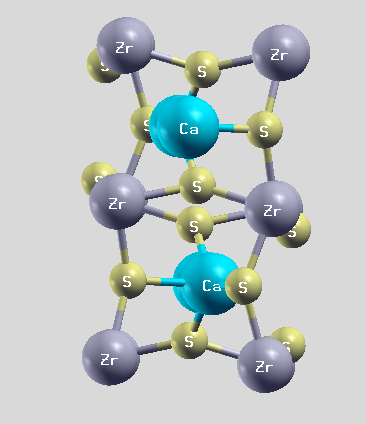}}
\caption{(Colour online) The GdFeO$_3$-type crystallographic structures of CaZrS$_3$.} 
\label{Fig1}
\end{figure}

Most of the material properties are governed by their crystal structures. A stable crystal structure is the one with the lowest energy 
arrangement of atoms at a given temperature and pressure~\cite{Sad2021}. In this study, an orthorhombic phase of CaZrS$_3$ [space group
$Pnma $(62) point group $D_{2h} (mmm)$] with the crystallographic structure of GdFeO$_3$-type was considered. Each Ca atom is centrally
located in the spatial region defined by its neighbouring S and Zr atoms in the lattice, which is composed of distorted ZrS$_6$ corner-sharing octahedral
as visualized in figure~\ref{Fig1}.  The ionic components are Ca$^{+2}$, Zr$^{+4}$ and S$^{-2}$ ions. 

The structural stability of perovskite materials in general is determined by the Goldschmidt tolerance factor $(\tau)$~\cite{Kum2021},
\begin{equation}
 \tau = \frac{1}{2}  \frac{(r_{\rm ca^+} + r_{\rm S^-})}{(r_{\rm Zr^+} + r_{\rm S^-})},
\end{equation}
where $r_{\rm Ca^+}, r_{\rm Zr^-}$ and $r_{\rm S^-} $ are ionic radii for  Ca$^{+2} $, Zr$^{+4}$ and S$^{-2} $  ions, respectively. The sizes of the 
ions are known to have a major influence on the structural distortion of perovskites. Thus, materials with a tolerance factor of 
$ 0.71\leqslant \tau  \leqslant 0.9 $  result in a distorted perovskite structure with tilted octahedral, while for
$ 0.9 < \tau < 1.0 $  the materials have an ideal cubic structure,  and when the tolerance factor is much higher ($\tau >1$) 
or lower ($\tau <0.71$), non-perovskite structures are commonly formed~\cite{Lel1980,Pil2020}. The ionic radii of Ca, Zr and S, 
and the calculated tolerance factor CaZrS$_3$ are shown in table~\ref{Tab1}. Here, since the calculated value falls within a 
distorted perovskite range, CaZrS$_3$ stability is defined, and \emph{Pnma} symmetry is adopted.

\begin{table}[htb]
\caption{The ionic radii and calculated tolerance factor of CaZrS$_3$.}
\label{Tab1}
\begin{center}
\begin{tabular}{ c c c }
\hline
Ions &  Radius (\AA) & Tolerance factor \\
\hline
Ca$^{+2}$    &   1.00  &       \\
Zr$^{+4}$   &   0.79  &   0.76 \\
S$^{-2}$    &   1.84  &        \\
\hline

\end{tabular}
\renewcommand{\arraystretch}{1}
\end{center}
\end{table}

Here, for structural optimization of CaZrS$_3$, the values of cutoff energy and $k$-point grid size obtained from the convergence test were utilized. 
Structural optimization was performed setting the convergence criteria; change in energy, $\Delta E =1.0\times 10^{-4}$~Ry and change in force, 
$\Delta F =1.0\times 10^{-3}$~Ry/\AA. The optimized equilibrium lattice constants were  $ a= 6.57$, $b=7.06$, and $ c=9.63$. These values are in good agreement with the 
theoretical and experimental values, as summarized in table~\ref{Tab2}. Moreover, a series of strained lattices were used to calculate the static 
lattice potential corresponding to total energy. From these results, the equilibrium unit cell volume, bulk modulus, and its pressure derivative 
can be calculated. A series of total energy calculations as a function of volume can be fitted to an equation of state (EOS) according to 
Murnaghan~\cite{Mur1944}:
\begin{equation}
 E(V)= E_0 + \frac{B_0 V}{B'_0} \bigg[ { \frac{(V_0 /V)}{B'_0} +1} \bigg] - \frac{B_0 V_0}{B'_0 -1},
\end{equation}
where $B_0$ is an equilibrium bulk modulus that effectively measures the curvature of the energy versus volume curve about
the relaxed volume $V_0$, and $B'_0$  is the derivative of the bulk modulus.

The calculated values
of the bulk modulus, equilibrium unit cell volume and the dimensionless bulk modulus derivative of CaZrS$_3$ are given in table~\ref{Tab2}. 
The calculated values of the unit cell volume and bulk modulus are in a good agreement with the experimental and the previous theoretical values,
respectively as shown in table~\ref{Tab2}.

\begin{table}[htb]
\caption{The calculated values of equilibrium lattice constant, unit cell volume, bulk modulus and its derivative of CaZrS$_3$ in comparison to existing works.}
\label{Tab2}
\vspace{1ex}
\begin{center}
\renewcommand{\arraystretch}{1.}
\begin{tabular}{ c  c c c c c c }
	\hline
\multirow{2}{*}{Source} & \multicolumn{3}{c}{Lattice parameter (\AA)} & \multirow{2}{*}{Vol (\AA$^3)$} & \multirow{2}{*}{$B$ (GPa)}  & \multirow{2}{*}{$B'$} \\
    &  a  &   b     & c  &  &  & \\
\hline
The calculated value  &   6.57 &  7.06   &  9.63  &  447.78  & 81.8  &  4.09    \\
\hline
Theory~\cite{Oum2019}$^a$        &     7 .0856 & 9.6647 & 6.5588 & 449.1479 & 82.4513 & 4.1716     \\
\hspace{1.3cm}~\cite{Oum2019}$^b$  &     7.0719 & 9.6611 & 6.5817 & 449.6771 & 105.7320 & $-$    \\
\hspace{1.3cm}~\cite{Kum2021}$^c$  &     6.56  & 7.06  & 9.63  &    &     &     \\
\hspace{1.3cm}~\cite{Zha2017}$^c$  &     7.02  & 6.47  & 9.53  &    &     &     \\
\hspace{1.3cm}~\cite{Maj2020}$^c$  &     7.07  & 9.63  & 6.57  &    &     &     \\
\hline
Experimental~\cite{Gla1972} &     7.03  & 6.54  & 9.59  &    &     &     \\
\hspace{1.75cm}~\cite{Lel1980}  &     7.03  & 9.59  & 6.54  & 440.66   &     &     \\
\hline
\multicolumn{7}{c}{\small $^a$(FP-LAPW method) DFT implemented in WIEN2K.} \\ 
\multicolumn{7}{c}{\small $^b$(PP-PW method) DFT implemented in WIEN2K.} \\
\multicolumn{7}{c}{\small $^c$(PAW method) DFT implemented in VASP.} \\
\hline

\end{tabular}
\renewcommand{\arraystretch}{1}
\end{center}

\end{table}


\subsection{Elastic properties}

\subsubsection*{Mechanical properties}
The elastic constant of crystals gives fundamental information for the study of mechanical characteristics of materials, as they are related to the mechanical properties of the material such as the elastic moduli, Poisson's ratio, and elastic anisotropy factor of materials. To calculate the elastic constants, we applied the non-volume-conserving method. The complete elastic constant tensor was determined from calculations of the stresses induced by small deformations of the equilibrium primitive cell. The elastic constant tensors $C_{ijk}$   are given by~\cite{Kar1997,Wal1972, Lua2018};
\begin{equation}
 C_{ijkl}=\left.\frac{\partial\sigma_{ij}}{\partial\varepsilon_{kl}}\right|_{\chi}=\frac{1}{V} \left.\frac{\partial^2 E}{\partial\varepsilon_{ij} \partial{\varepsilon_{kl}}} \right|_\chi ,
\end{equation}
where $E$  stands for the Helmholtz free energy, $\sigma_{ij}$  and $\varepsilon_{kl}$ are the applied stress and Eulerian strain tensors, and $\chi$ stands for the coordinates.

In this case, for  orthorhombic system, there are nine independent elastic constants that should satisfy the well-known Born stability criteria~\cite{Voigt}.
\begin{eqnarray}
 &&(C_{11}+C_{22}-2C_{12})>0, \quad (C_{11} + C_{33}-2C_{13})>0, \nonumber  \\
&&(C_{22}+C_{33} -2C_{23})>0, \quad C_{11}>0 , \quad C_{22}>0, \nonumber \\
 &&C_{33} >0 ,\quad C_{44} > 0,\quad C_{55} >0,\quad C_{66} >0, \nonumber \\
&&(C_{11} + C_{22}+ C_{33}+ 2C_{12}+2C_{13}+2C_{23})>0. 
\end{eqnarray}
The calculated elastic constants (table~\ref{Tab3}) satisfy the mechanical stability conditions above and $C_{ij} >0$, $C_{12} < C_{11}$ and $C_{13} < 1/2 (C_{11} + C_{33}) $.

\begin{table}[htb]
	\caption{The calculated elastic constants of CaZrS$_3$.}
	\label{Tab3}
	\begin{center}
		\begin{tabular}{ c  c    c c c c c c c c }
			\hline
			& $C_{11}$ &  $C_{12}$ &  $C_{13}$ &   $C_{22}$ &  $C_{23}$ &     $C_{33}$ &  $C_{44}$ &  $C_{55}$ &  $C_{66}$  \\
			\hline
			The calculated &  \multirow{2}{*}{120.9} & \multirow{2}{*}{59.4} & \multirow{2}{*}{38.5}  & \multirow{2}{*}{155.47} & \multirow{2}{*}{36.91} &  \multirow{2}{*}{146.5} & \multirow{2}{*}{44.5} &  \multirow{2}{*}{28.6} &    \multirow{2}{*}{51.4 }   \\ value    \\
			\hline
			
		\end{tabular}
		\renewcommand{\arraystretch}{1}
	\end{center}
	
\end{table}

Using the Voigt-Reuss-Hill (VRH) average approximation, mechanical parameters such as the Young modulus ($E$), Poisson's ratio ($\eta$) and shear modulus
($G$) are computed from the calculated elastic constants as~\cite{Voigt,Wal1981}, the elastic constants and calculated mechanical properties from the 
elastic constants are shown in table~\ref{Tab3} and table~\ref{Tab4}, respectively.


\begin{table}[htb]
\caption{Mechanical properties calculated from elastic constants of CaZrS$_3$.}
\label{Tab4}
\begin{center}
\begin{tabular}{ c  c c c c c c c c c c  }
\hline
     &   $B$ &  $G $ &  $ B/G $ &  $E $ &  $\eta $ &  $A^u$   \\
\hline
The calculated  &   \multirow{2}{*}{76.36} & \multirow{2}{*}{42.73} & \multirow{2}{*}{1.78} &  \multirow{2}{*}{108.04} & \multirow{2}{*}{0.26} & \multirow{2}{*}{0.35}  \\ value   \\
\hline

\end{tabular}
\renewcommand{\arraystretch}{1}
\end{center}

\end{table}

Bulk modulus $B$ is an essential physical parameter in describing the compressibility of solids under the hydrostatic pressure. Large value
of bulk modulus results in higher compressibility of a solid material. 
The Bulk modulus value (76.36~GPa) calculated from elastic constants is found to be close to the value calculated from the equation of state (81.8~GPa) 
as in  table~\ref{Tab2}. The shear modulus $G$ is another important parameter that can describe the shape change under the shear force. The larger the 
shear modulus is, the higher is the shape change resistance of the solid  material. The calculated value of the shear modulus is 42.73~GPa and it is
comparable to the value obtained by~\cite{Oum2019}. Furthermore, determining the ratio of $B/G$ is important for understanding
the brittle and ductile behavior of materials in the material                                                   
fabrication. The ductility and brittleness of materials can be determined based on the value of $B/G$ ratio according to Pugh~\cite{Pug1954}.
The cutoff value is 1.75. When $B/G > 1.75$, the material behaves in a ductile manner, otherwise, it exhibits brittle properties.
From table~\ref{Tab4}, our calculated ratio for $B/G$ value is 1.78 for CaZrS$_3$ in GdFeO$_3$-type phase, showing that CaZrS$_3$ in 
this phase is ductile. Another mechanical parameter that provides information about the feature of the bonding forces is Poisson's ratio ($\eta$).  
In the evaluation of Poisson's ratio, the values  0.25 and 0.5 are the lower and upper limits of the central force, respectively~\cite{Rav1998}. 
From table~\ref{Tab4}, the calculated Poisson's ratio ($\eta$) is 0.26 (which is between 0.25 and~0.5) indicating that the inter-atomic forces are central.

The universal anisotropic index [$ A^u = ({5G_v}/{G_R}) + ({B_v}/{B_R}) -6$] is a measure to define the elastic anisotropic or isotropic characteristics based on the contributions of 
both bulk and shear modulus~\cite{Ran2008}. It is one of the important physical parameters used to study the service life time of materials.
 The material is isotropic if the value of $ A^u = 0 $; otherwise it ($ A^u \neq 0 $) refers to the anisotropic mechanical properties. Any value smaller 
 or greater than zero represents a higher extent of anisotropy. Based on this, for orthorhombic phase of CaZrS$_3$,  the calculated value for $A^u $ is 0.35,
 indicating that the CaZrS$_3$ was found to be anisotropic. 
                                                             
\subsubsection*{Debye temperature}       
The thermal properties of a solid material are related to two physical parameters: Debye temperature~($\theta_{\rm D}$) and melting temperature $M_t$, 
respectively. The Debye temperature is another essential physical term that can be used to characterize solid-state physics phenomena such as 
lattice vibration, elastic constants, specific heat, and melting point. The magnitude of the Debye temperature is helpful to know the thermal conductivity
of solid materials. The higher the value of the Debye temperature is, the higher is its thermal conductivity. The Debye temperature ($\theta_{\rm D}$) of CaZrS$_3$ can 
be estimated from 
the averaged sound velocity, $C_m$, given by~\cite{Jia2010}, 
\begin{equation}
	\theta_{\rm D}=\frac{h}{k_{\rm B}} \bigg[\frac{3n}{4\piup}\bigg(\frac{N_{\rm A}\rho}{M}\bigg)	\bigg]^{{1}/{3}} C_m , \qquad    
	C_m= \bigg[ \frac{1}{3} \bigg(\frac{2}{c_t^2} + \frac{1}{c_l^3} \bigg) \bigg]^{{-1}/{3}},
\end{equation}
where $h$ is Plank's constant, $k_{\rm B}$ is Boltzmann's constant, $N_{\rm A}$ is Avogadro's number, $\rho$  is density, $M$ is molecular weight, $n$ is the number of 
atoms in a formula unit, $c_l$ is the longitudinal sound velocity,  and $c_t$ is the transverse sound velocity.  The longitudinal and transverse sound velocities can be obtained from density, shear and bulk modulus of the material as:
\begin{equation}
 c_l= \bigg( \frac{B+\frac{3}{4}G}{\rho}\bigg)^{{1}/{2}}, \qquad c_t=
 \bigg( \frac{G}{\rho}\bigg)^{{1}/{2}}.
\end{equation}
Moreover, the Debye average sound velocity which represents the maximum
frequency of the material is described by $c_{\rm D}= \big( {k_{\rm B}	T}/{\rho}\big)^{{1}/{2}}$. The melting point of a material depends on Debye temperature; a larger Debye temperature of the material shows a higher melting temperature~\cite{Lu2017}.
The melting temperature $M_t$  of a stable phase of CaZrS$_3$ can be determined based on an elastic constant $C_{11} $ using~\cite{Fin1984},
\begin{equation}
 M_t= 553\, {\rm K}+5.9C_{11}\, {\rm K}.
\end{equation}
 The calculated values of longitudinal sound velocity, transverse sound velocity and Debye temperature for CaZrS$_3$ are given in table~\ref{Tab5}. 
It was observed that the calculated values for Debye temperature and melting point are 415.5~K, and 1267.6~K, respectively.


\begin{table}[htb]
\caption{Density, sound velocities, Debye temperature and melting point of  CaZrS$_3$.}
\label{Tab5}
\begin{center}
\renewcommand{\arraystretch}{1.5}
\begin{tabular}{ c  c c c c c c c   }
\hline
     & $\rho$ (gcm$^{-3}$) &  $c_l$ (m/s) &  $ c_t$ (m/s)  &  $c_m$ (m/s)  &  $c_{\rm D}$ (m/s) &  $\theta_{\rm D}$ (K) &  $ M_t$ (K)   \\
  \hline                                                                                                       
The calculated value  &  3.4826 & 6283.72 & 3557.14 & 3759.0 & 3932.83 & 415.5 & 1267.6    \\
\hline

\end{tabular}
\renewcommand{\arraystretch}{1}
\end{center}

\end{table}

\subsection{Electronic properties: density of states (DOS) and band structure}

\begin{figure} [htb]
    \centering
    {\includegraphics[width=6.15 cm]{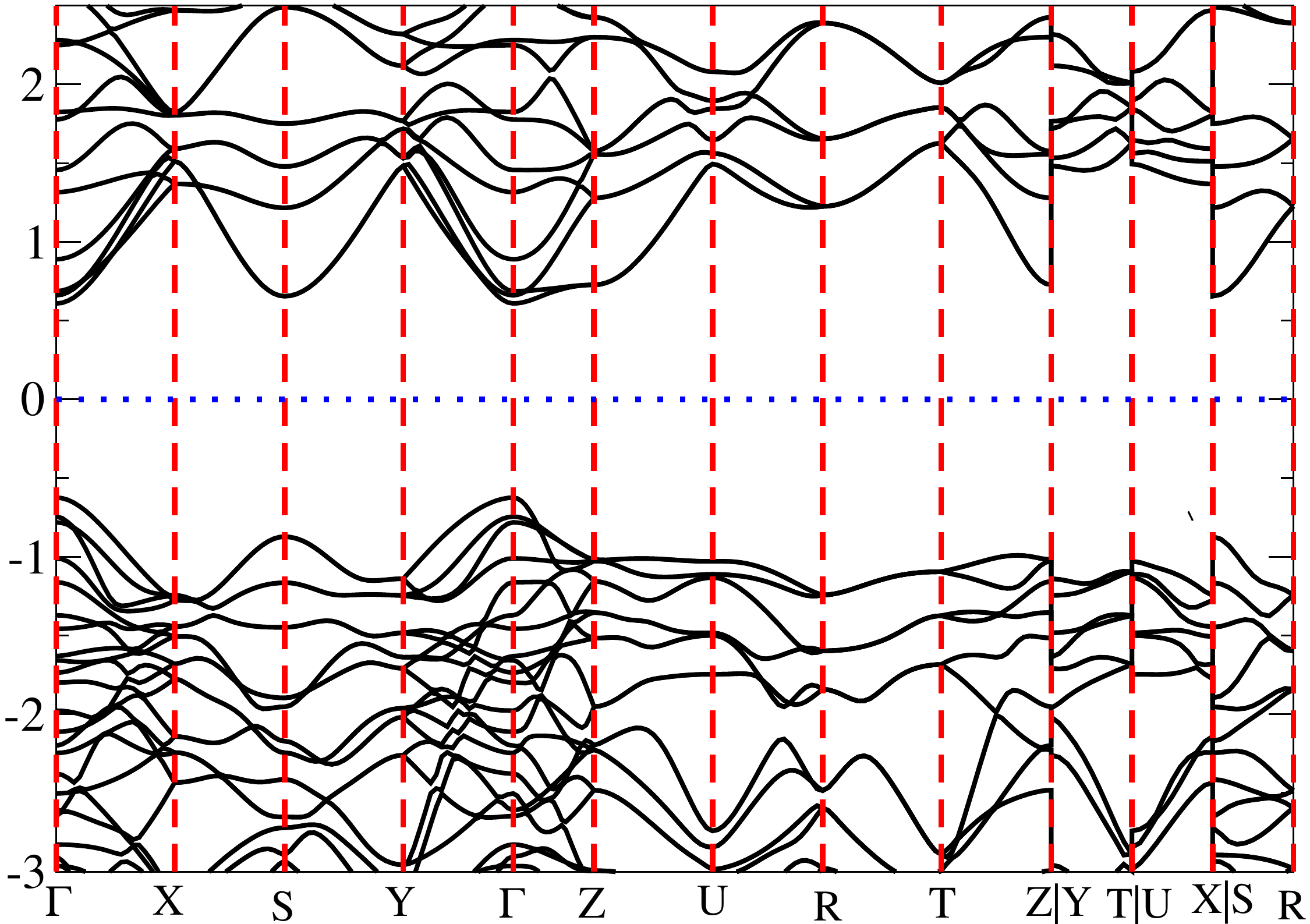} }%
    \qquad
{\includegraphics[width=6.15 cm]{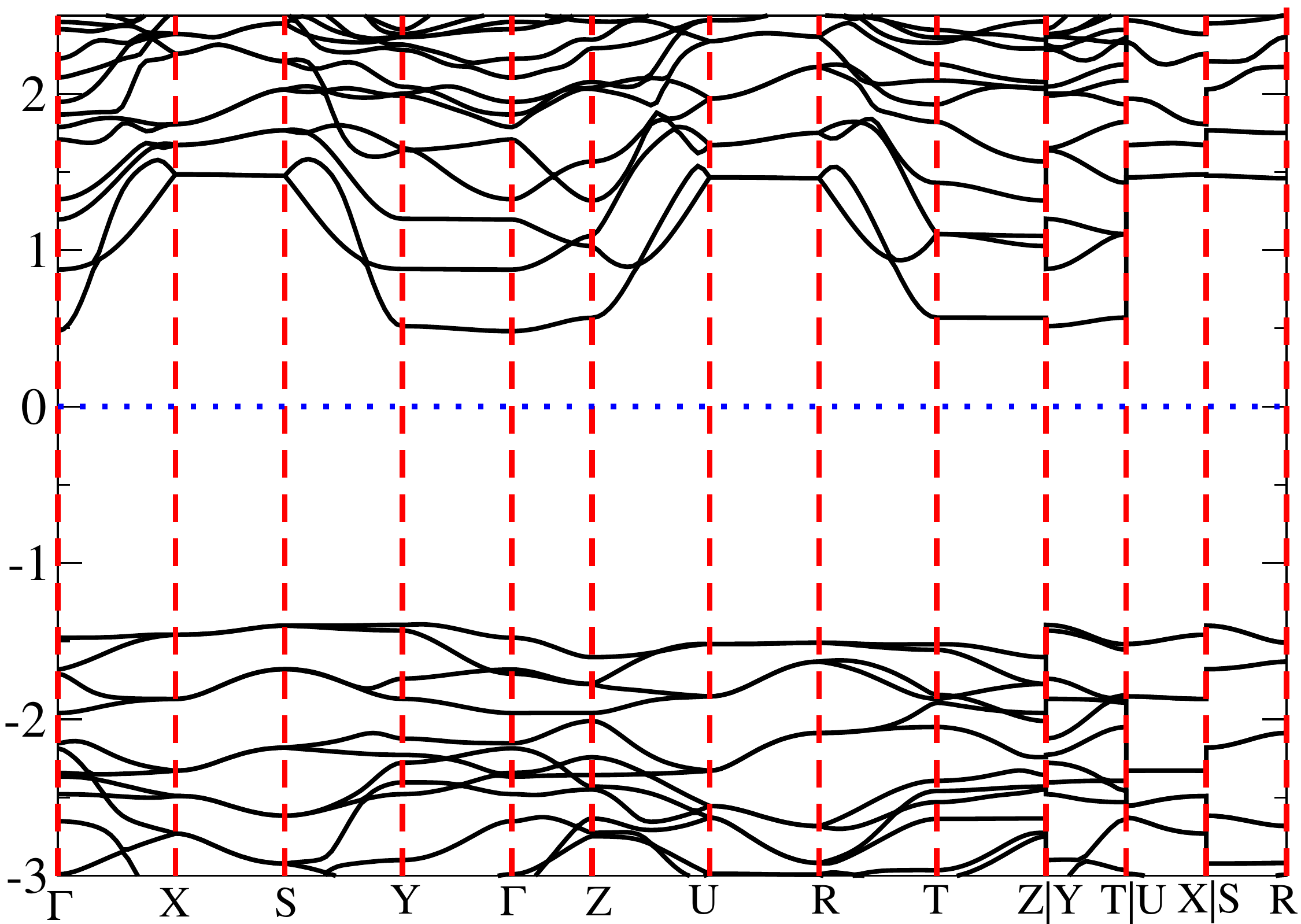} }%
    \caption{(Colour online) Band structure of CaZrS$_3$ with respect to GGA-PBE (left-hand) and GGA+U (right-hand).}%
    \label{Fig3}
\end{figure}

The energy band structure and density of states of materials are used to determine the electronic properties of solid materials. 
The accessible electronic energy levels of solid materials are represented by electronic band structures. In this study, the 
electronic band structure along the high symmetry direction of the Brillouin zone was estimated using the GGA-PBE functional 
and the Hubbard correction (GGA+U) for exchange correlation potential, as shown in figure~\ref{Fig3}. The GGA-PBE functional 
fails to approximate the exact exchange correlation potential, since the band gap value obtained by approximating the exchange 
correlation potential with GGA-PBE functional is 1.23~eV (table~\ref{Tab6}), which appears to be underestimated as compared to experimental band gap 
values of 1.90~eV~\cite{Per2016,Gla1972}. Furthermore, the band gap was calculated with the Hubbard correction for on-site interaction, 
yielding a band gap of 1.88 eV, which is close to the experimental result~\cite{Per2016,Gla1972}.	

\begin{table}[htb]
\caption{The calculated band gap value of CaZrS$_3$ in comparison to the existing theoretical and experimental results.  }
\label{Tab6}
\begin{center}
\begin{tabular}{ c  c c c    }
\hline
       Source &   GGA-PBE & HSE06 & GGA+U  \\
\hline
The calculated value  &    1.23 & $-$ & 1.88     \\
\hline
Theory~\cite{Kum2021}      &      1.24 &    2.04 &      \\
\hspace{1.3cm}~\cite{Zha2017}   & &  2.22&    \\
\hline
Experimental~\cite{Per2016,Gla1972} &    1.90 &  &       \\
\hline
\end{tabular}
\renewcommand{\arraystretch}{1}
\end{center}

\end{table}

\begin{figure} [htb]
    \centering
    {\includegraphics[width=5.75cm]{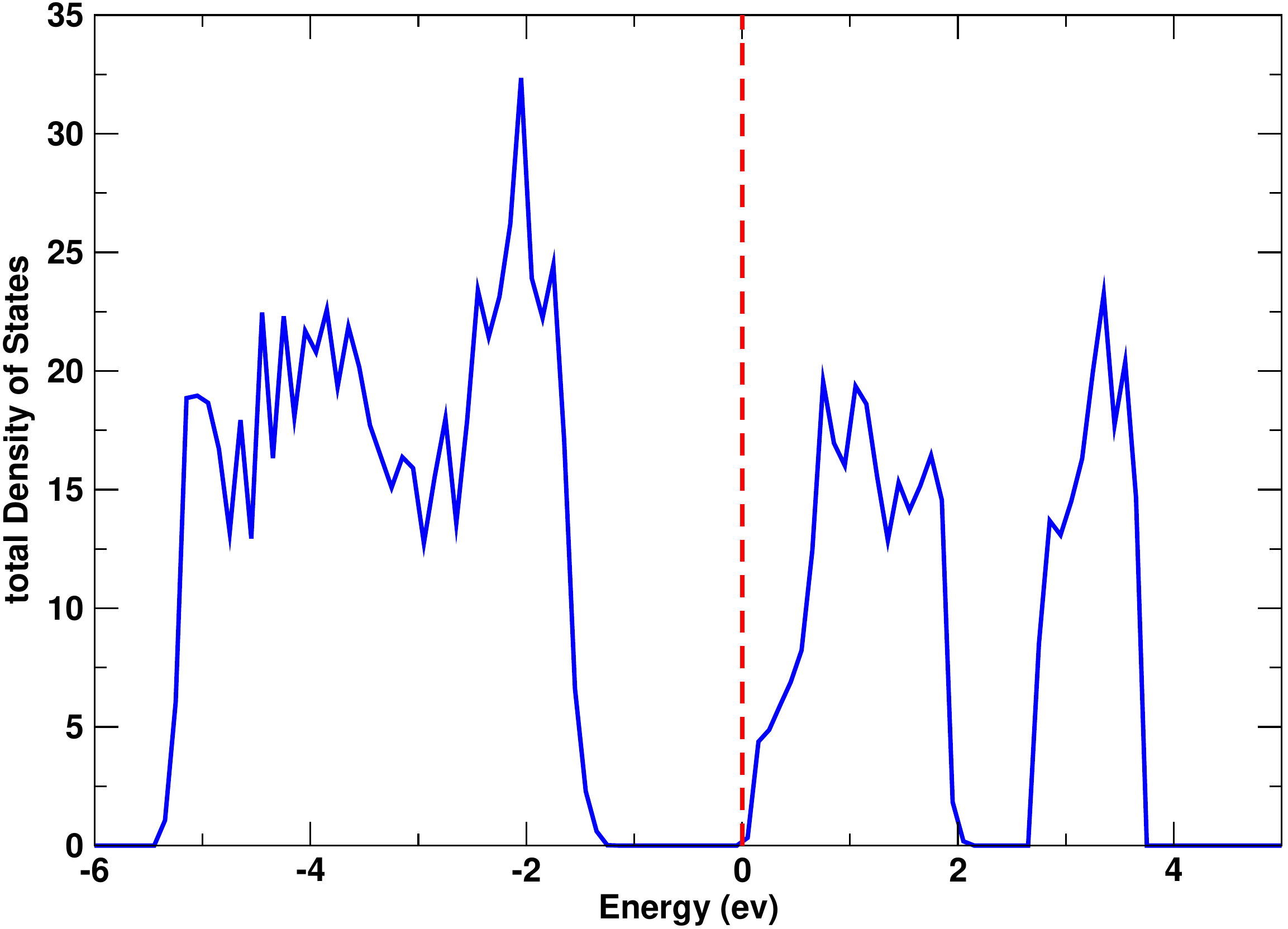} }%
    \qquad
    {\includegraphics[width=5.75cm]{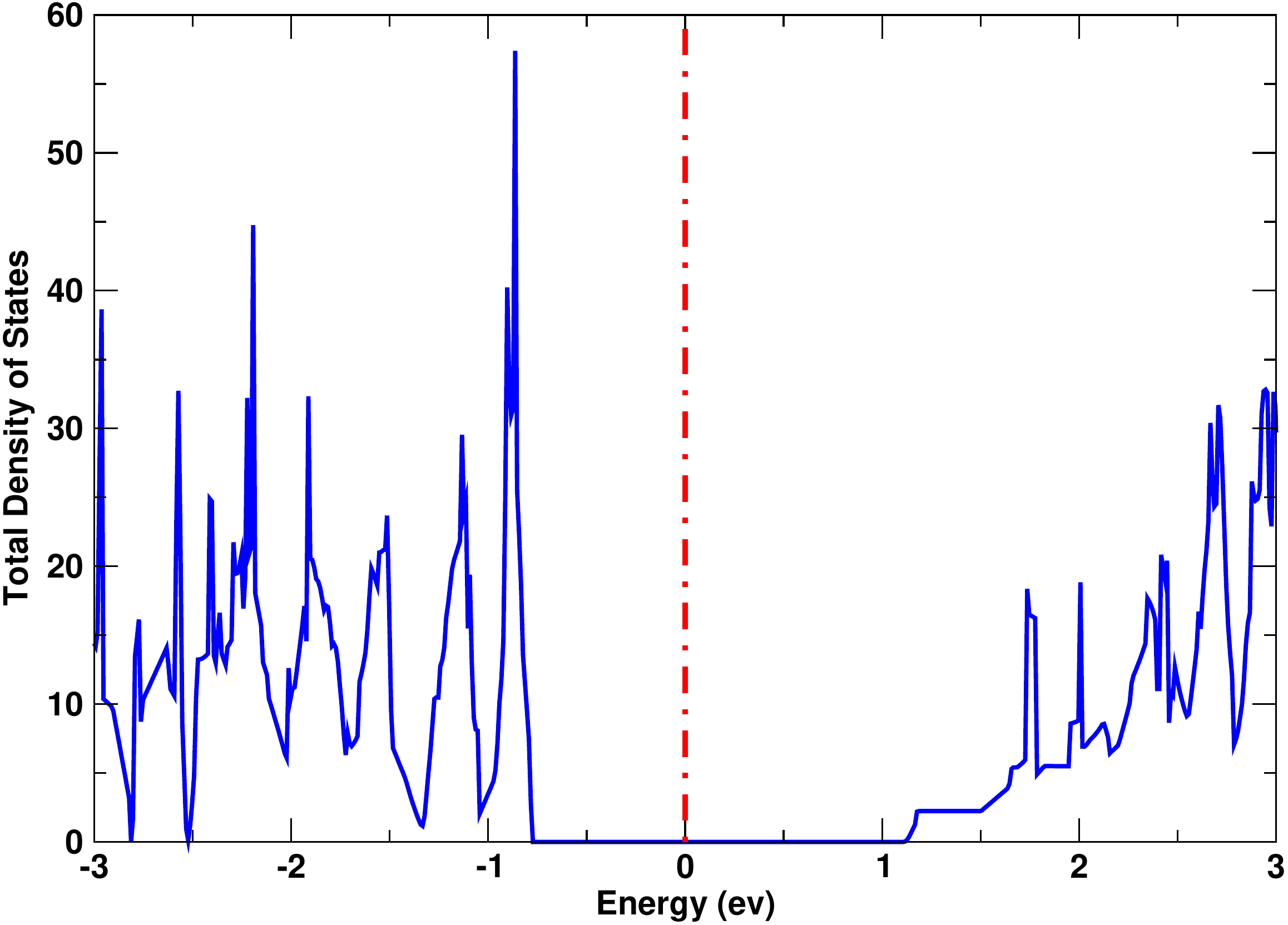} }%
    \caption{(Colour online) Total density of states of CaZrS$_3$ with respect to GGA-PBE (left-hand) and GGA+U (right-hand).}%
    \label{Fig4}
\end{figure}

\begin{figure} [htb]
	\centering
	{\includegraphics[width=5.75cm]{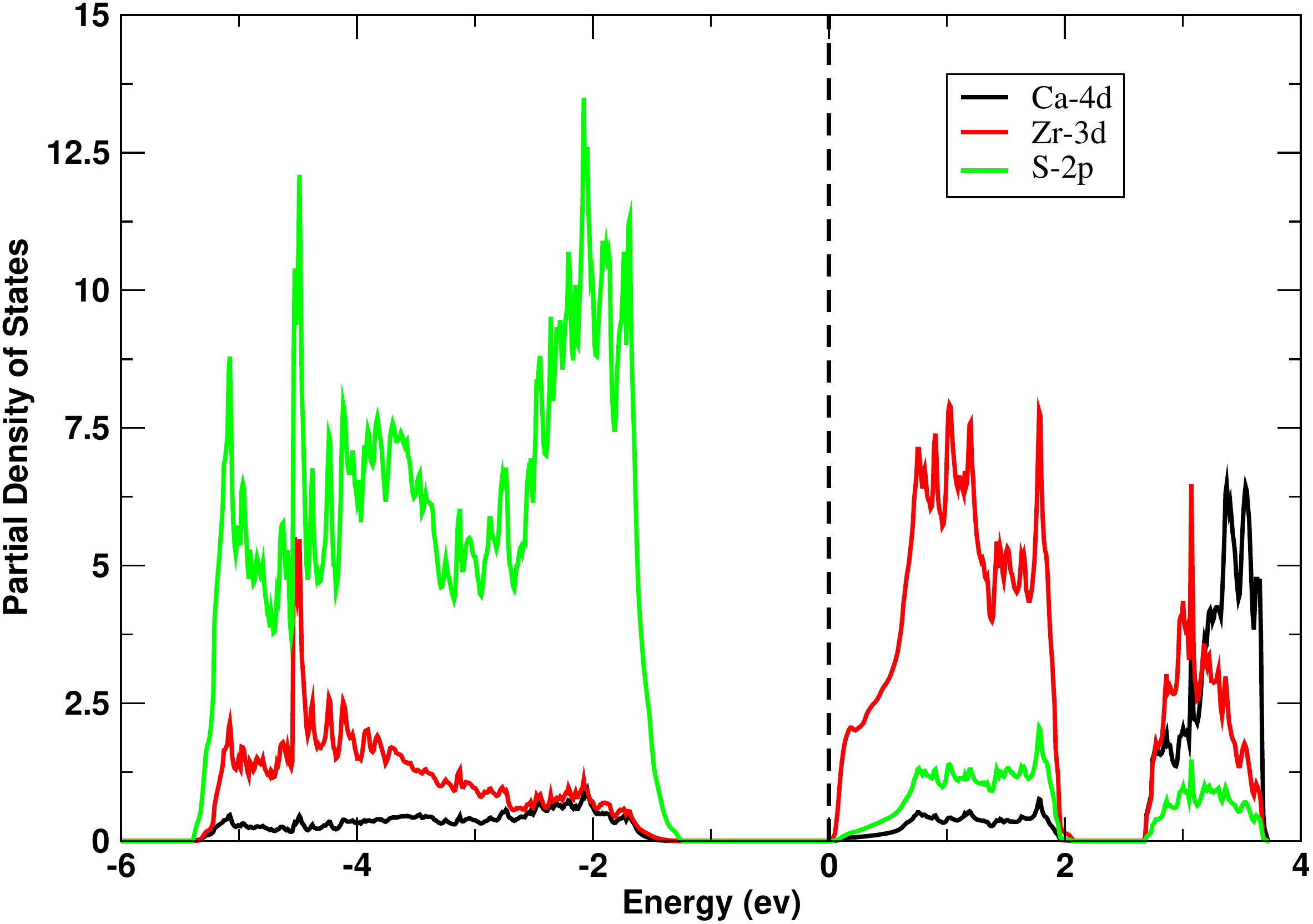} }%
	\qquad
	{\includegraphics[width=5.75cm]{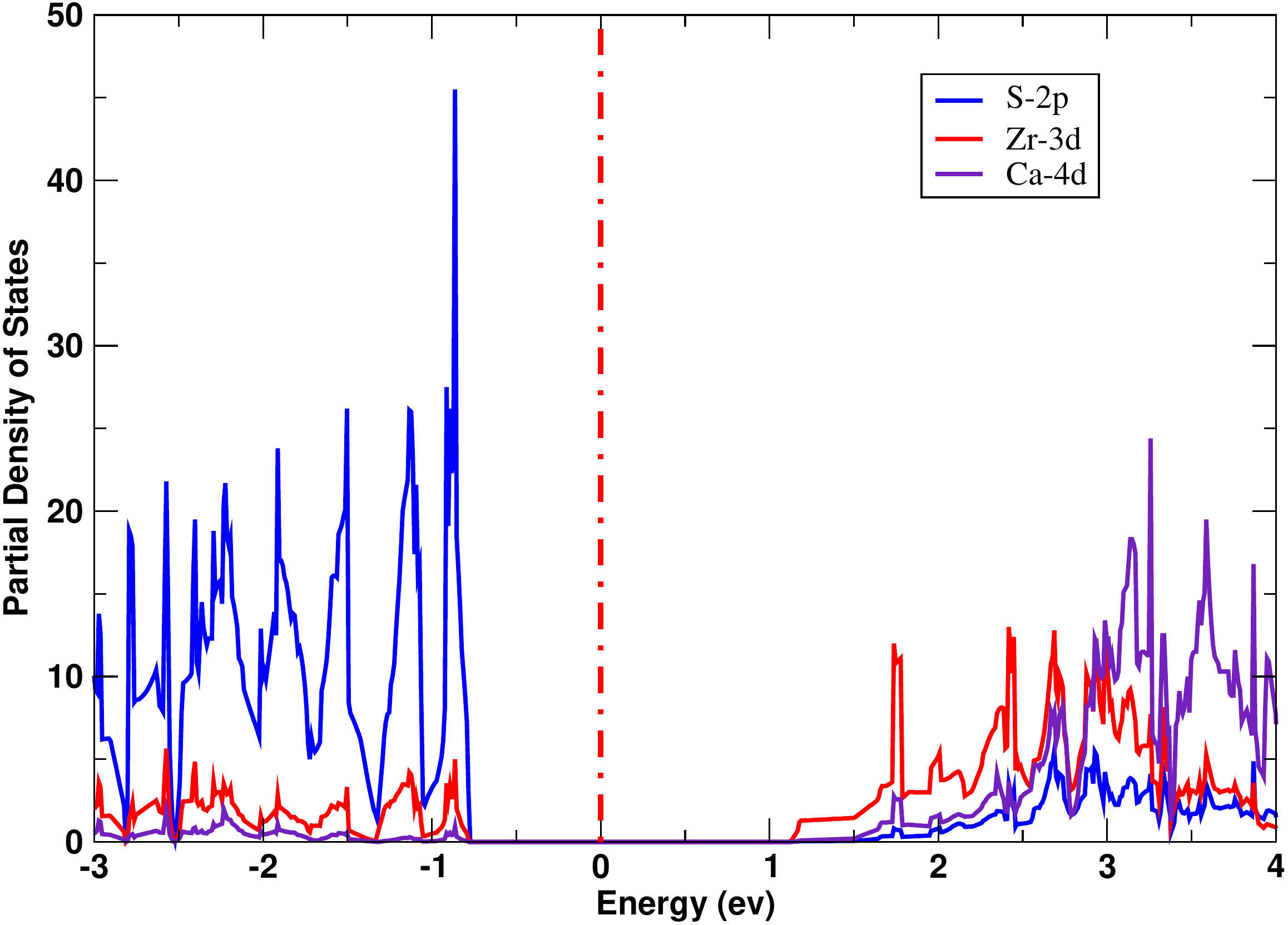} }%
	\caption{(Colour online) The partial densities of states of CaZrS$_3$ with respect to GGA-PBE (left) and GGA+U (right).}%
	\label{Fig5}
\end{figure}

The DOS is also used to describe how state occupancy behaves at different energy levels. It provides information on
both occupied and empty states. The states that are available for occupancy have a high DOS at a certain energy level. However, there is 
no state occupied at DOS equal to zero. In this study, the total and partial densities of states were obtained for the equilibrium states 
of the phases using GGA-PBE correlation interaction and also with GGA+U as shown in figure~\ref{Fig4} and figure~\ref{Fig5}. 
The density of states is discontinuous for the width from the top of the valence band to the bottom of the conduction band 
which is normally refered to the band gap of the system. Moreover, figure~\ref{Fig5} shows that the maximum valence band is mainly contributed by S-$2p$ orbitals 
and the minimum conduction band is mainly dominated by Zr-$3d$ orbitals. On the other hand, Ca-$4d$ orbitals are observed on both maximum valence and minimum 
conduction bands, and the rest of the orbitals have a small contribution. 

\subsection{Phonon dispersion relation}

Phonon vibration plays an essential role in dynamic behaviors and in thermal properties, which are central topics in fundamental issues of materials science.
The phonon frequency of crystalline structures is one of the fundamental aspects when considering the phase stability, phase transformations, 
and thermodynamics of these materials.  The phonon density of states $g(\omega)$ is given by~\cite{Cha2005,Dov1993}
\begin{equation}
 g(\omega)=\frac{1}{N} \sum_{qj} \delta(\omega-\omega_{qj}),
\end{equation}
where $N$ is the number of unit cells in a crystal. Divided by $N$, $g(\omega)$ is normalized so that the integral over frequency becomes $3n_a$, where
$n_a$ is the number of atoms.

Considering the atom specific phonon density of states projected along a unit direction vector $\hat{n}$,  is defined as~\cite{Dov1993}
\begin{equation}
 g_{k} (\omega, \hbar)=\frac{1}{N} \sum_{qj} \delta(\omega-\omega_{qj}) \big|\hbar\re^k_{qj}\big|^2.
\end{equation}
From the canonical distribution in statistical mechanics for phonons under the harmonic approximation, the energy $E_n $ of the phonon system is given as 
\begin{equation}
 E=\sum \hbar\omega_{qj} \bigg[ \frac{1}{2} + \frac{1}{\exp(\hbar \omega_{qj} /k_{\rm B} T) -1}\bigg],
\end{equation}
where $T$, $k_{\rm B}$  and $\hbar$ are the temperature, the Boltzmann constant, and the reduced Planck constant, respectively. Here, from statistical
mechanics, $\hat{n}={1}/[{\exp(\hbar \omega_{qj} /k_{\rm B} T) -1}] $  gives the mean phonon number distribution function.

\begin{figure}[htb]
	\centerline{\includegraphics[width=0.6\textwidth]{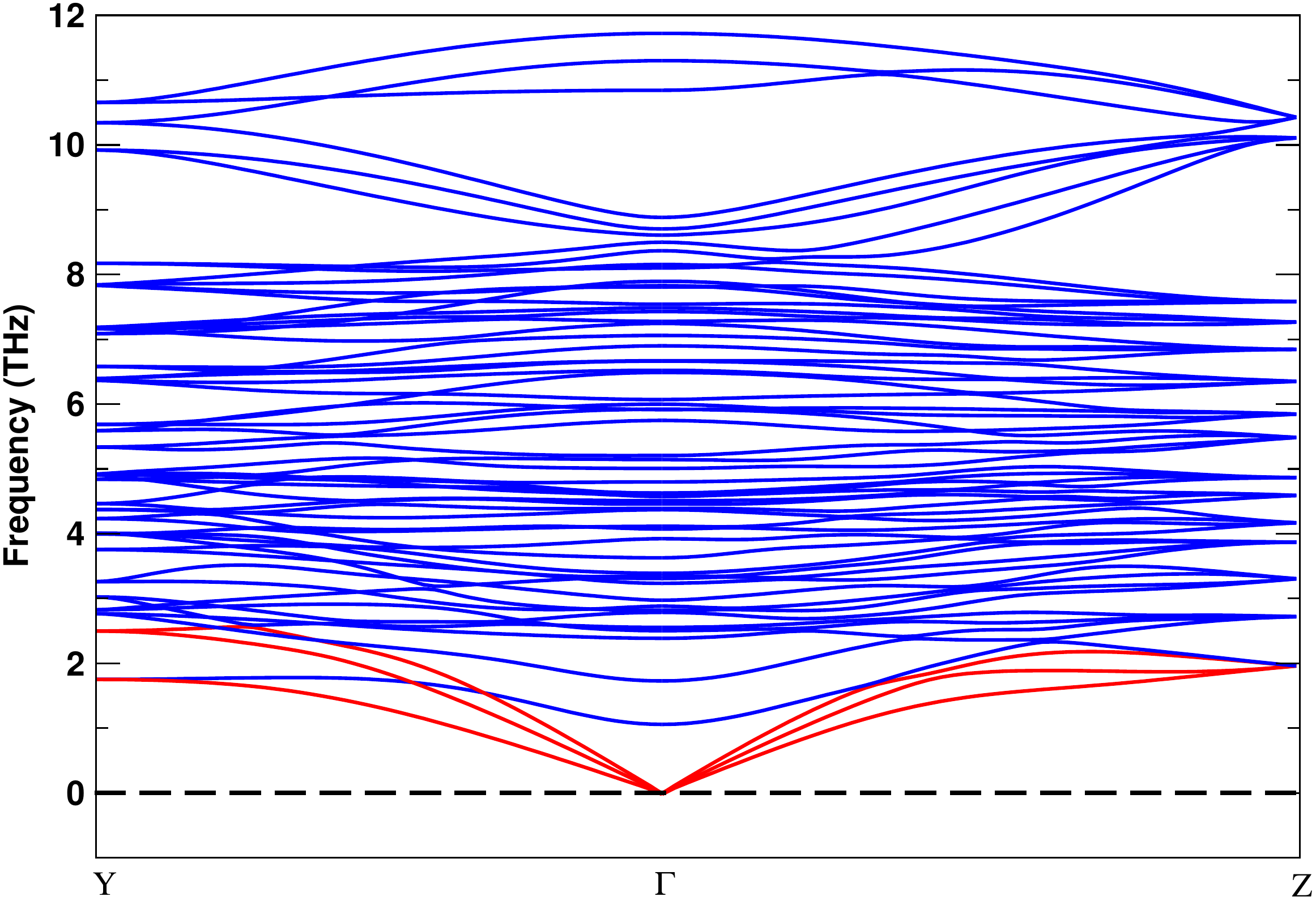}}
	\caption{(Colour online) Phonon dispersion relation for CaZrS$_3$.} 
	\label{Fig6}
\end{figure}

\begin{figure}[htb]
	\centerline{\includegraphics[width=0.6\textwidth]{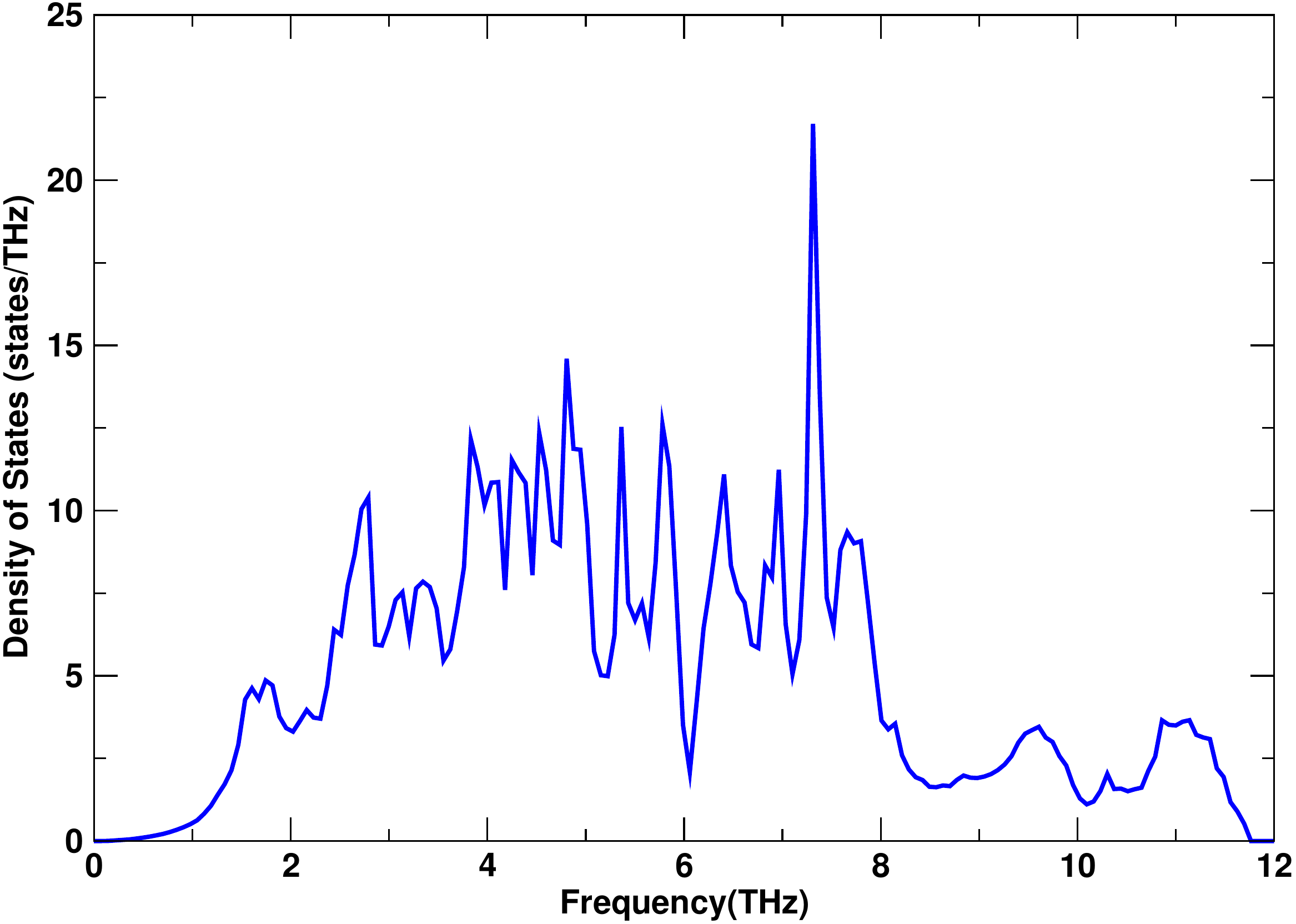}}
	\caption{(Colour online) Phonon density of states of CaZrS$_3$.} 
	\label{Fig7}
\end{figure}

A supercell of $1 \times 1 \times 2$ (in $x$, $y$, and $z$-direction) containing 40 atoms was created to study the phonon dispersion relation for  CaZrS$_3$, using a
PHONOPY package with Quantum ESPRESSO package as implemented in~\cite{Tog2015}. It is known that a crystal constituent of 40 atoms in bulk 
system (three dimensions~$x$, $y$, $z$ coordinates) has 120 degrees of freedom ($3n_a$, where $n_a$ is the number of atoms). The phonon dispersion relation 
for frequency bands and frequency density of states were calculated and displayed as shown in figure~\ref{Fig6} and figure~\ref{Fig7}, respectively. 
As indicated in figure~\ref{Fig6}, it was observed that there are three (3) acoustic branches and 117 optical branches ($3n_a -3 $) mode of vibrations.
Here, also the results showed that CaZrS$_3$ possesses no imaginary phonon frequency modes. Hence, it is structurally and lattice dynamically stable.
 This finding agrees with the results of the analysis of the elastic constants and Goldschmidt tolerance factor.

\subsection{Thermodynamic properties}

\begin{figure}[htb]
\centerline{\includegraphics[width=0.85\textwidth]{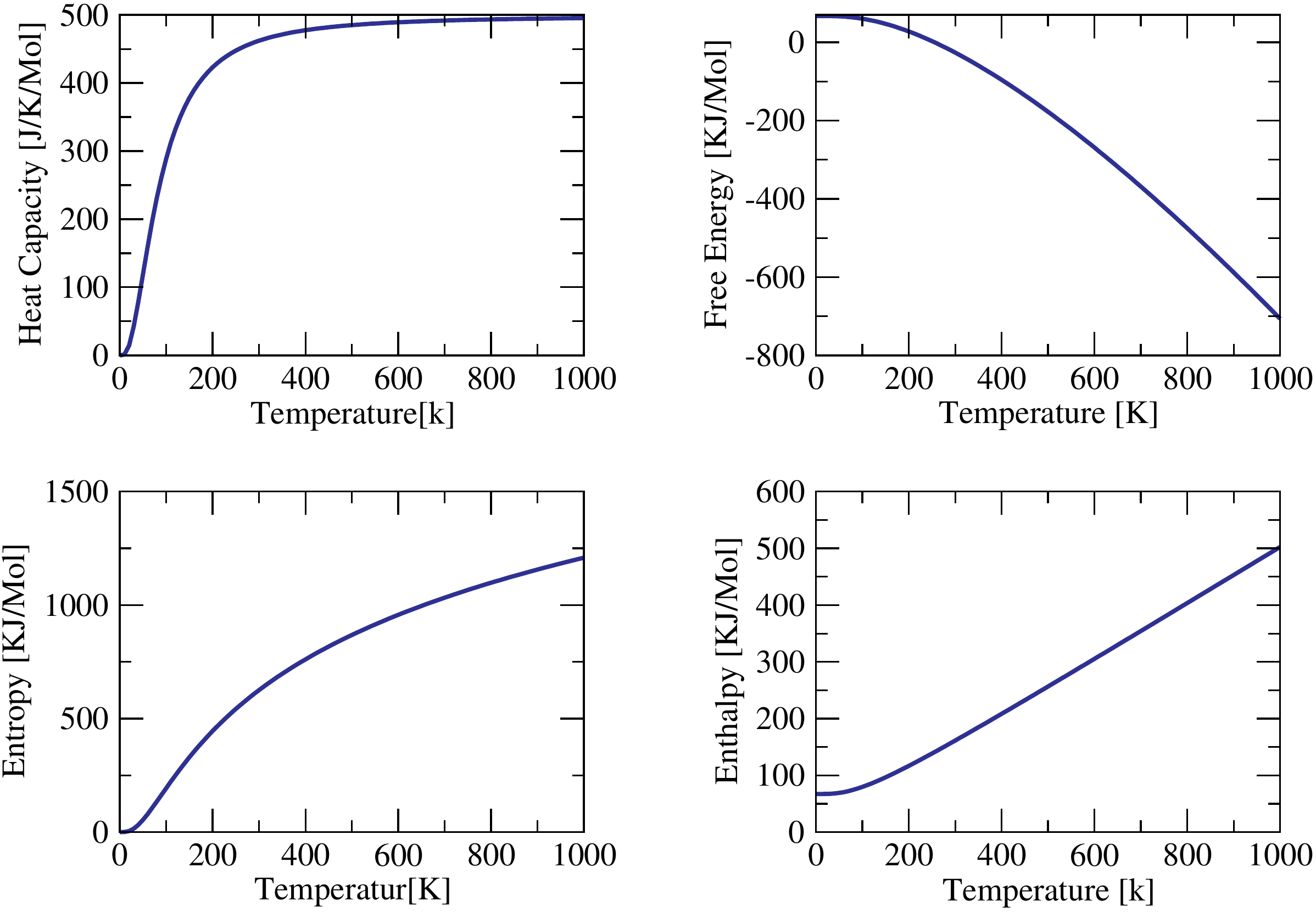}}
\caption{(Colour online) Specific heat, free energy, entropy, and enthalpy  of CaZrS$_3$ with respect to temperature variation. } 
\label{Fig8}
\end{figure}

Thermodynamic properties of materials are one of the foundations of solid-state science and industry. The investigation of these properties is
important in order to determine their specific behavior when these materials are subjected to high pressure and temperature. Using the thermodynamic 
relations, a number of thermal properties, such as constant volume heat capacity $C_v$, Helmholtz free energy $F$, and entropy~$S$, can be computed as functions 
of temperature as~\cite{Wal1972,Cha2005, Gro2017} 
\begin{equation}
 C_V= \sum_{qj} k_{\rm B} \left( \frac{\hbar\omega_{qj}}{k_{\rm B} T}\right)^2 \frac{\exp\big(\hbar\omega_{qj}/k_{\rm B} T\big)}{\big[\exp\big(\hbar\omega_{qj}/k_{\rm B} T\big)-1  \big]^2},
\end{equation}
\begin{equation}
 F= \frac{1}{2} \sum_{qj} \hbar \omega_{qj} +  k_{\rm B} T \sum_{qj} \ln\bigg[1-\exp\big(-\hbar \omega_{qj}/ k_{\rm B} T\big)  \bigg],
\end{equation}
\begin{equation}
 S=\frac{1}{2T} \sum_{qj} \hbar \omega_{qj} \coth\big(\hbar\omega_{qj}/2k_{\rm B} T \big) -k_{\rm B} \sum_{qj} \ln \big[ 2\sinh \big(\hbar\omega_{qj}/2k_{\rm B} T\big)  \big]. 
\end{equation}
Here also, a supercell of of $1 \times 1 \times 2$ (in $x$,  $y$, and $z$-direction) containing 40 atoms was used to study the thermodynamic properties of CaZrS$_3$, using
a PHONOPY package with Quantum ESPRESSO package as implemented in~\cite{Tog2015}. At finite temperatures ranging from 0~K to 1000~K, thermodynamic
parameters such as enthalpy, free energy, entropy, and heat capacity were computed and plotted in figure~\ref{Fig8}. In this consideration, the volume 
and temperature are independent variables. From figure~\ref{Fig8}, one can observe that below 10 K, the values of the entropy and heat capacity are 
almost zero. The free energy diminishes gradually as the temperature rises, whereas the entropy rises rapidly, following reasonable trends shown 
in~\cite{Cha2005,Lee1995,Foi1994}.  As a result, the enthalpy increases linearly with the increment of temperature. The increase in enthalpy at 
high temperatures leads to a decrease in the free energy which is associated with defects. It is clearly observed that for the temperature below 400~K, 
the heat capacity increases rapidly, whereas for temperature  above 400~K it increases slowly (almost increasing linearly) with temperature and 
gradually approaches the Dulong-Petit limit (classical limit)  owing to the anharmonic approximations of the Debye model as observed
in~\cite{Cha2005,Ose2020, Li2011}. The calculated heat capacity graph is also smooth and continuous confirming that there is no phase change 
occurring in CaZrS$_3$ up to 1000~K~\cite{Cha2005}.

\section{Conclusion}

Chalcogenide perovskites have emerged as a non-toxic and stable photovoltaic material that has similar optoelectronic capabilities to lead halide hybrid perovskites. In this study, the first-principles calculations using the Quantum ESPRESSO software package were used to study the structural, electrical, elastic, phonon dispersion relation, and temperature-dependent thermodynamic characteristics of orthorhombic CaZrS$_3$ for optoelectronic applications. The THERMO\_PW package was used to compute the elastic characteristics of the material using the optimized structure. The PHONOPY package was also used to calculate the phonon dispersion and thermal characteristics. The computed lattice parameters $ a= 6.57$, $b=7.06$, and $c= 9.63$ correspond well to theoretical and experimental results. The elastic constants were used to calculate mechanical parameters of CaZrS$_3$, such as the bulk modulus, shear modulus, Young's modulus, and elastic anisotropy. CaZrS$_3$ was found to be classified as a ductile material according to the determined value of  $B/G$ ratio (1.78). Poisson's ratio obtained the value of 0.26, indicating that interatomic forces are central. The measured global anisotropic index value of 0.35 validates the anisotropic nature of CaZrS$_3$. The band gap value of CaZrS$_3$ was calculated by approximating the exchange correlation potential with GGA/PBE and GGA+U. The band gap value calculated using GGA/PBE is 1.23~eV, which is $35\%$ percent less than the experimental result. However, the computed band gap value using GGA+U is 1.88~eV, which is close to the experimental band gap value. The absence of imaginary (negative frequencies in the figures) frequencies in phonon dispersion curve and the phonon density of states give an indication that the structure is dynamically stable. The temperature dependence of thermodynamic parameters including enthalpy, entropy, free energy, and heat capacity is calculated and analyzed.

\section*{Funding}

Funding: This study was funded by College of Natural Sciences for supporting PhD student (grant number: CNSPHYS-04-2021/2022), and also supported by subsidy fund by Arba Minch University (under project: Gov/AMU/TH3/CNS/phy/01/14).

\section*{Acknowledgement}
One of the authors, Mulugetta Duressa Kassa, expresses his thanks and appreciation to the Department of Physics, College of Natural Sciences, Jimma University, for providing financial suport for this study and Prof. Omololu Akin-Ojo (director, ICTP-EAIFR) for valuable suggestions on computational techniques.  The authors would also like to acknowledge EAFIR and ICTP for High Performance Cluster Machine (HPC) support during the computation.




\ukrainianpart

\title[Дослідження структурних, пружних, електронних, фононних дисперсійних і термодинамічних властивостей]{Дослідження структурних, пружних, електронних, фононних дисперсійних і термодинамічних властивостей орторомбічного CaZrS$_3$ для застосувань в оптичній електроніці}
\author{М. Д. Касса,  Н. Г. Дебело, М. М. Волдемаріям} 
\address{Фізичний факультет університету Джимми, 378, Джимма, Ефіопія}

\makeukrtitle

\begin{abstract}
	\tolerance=3000%
	
Халькогенідні перовскіти забезпечують відмінну термічну та водну стійкість, а також доброякісний елементний склад  для оптоелектронних застосувань у порівнянні з органічними галогенідними перовскітами.
У даній роботі структурні, електричні, пружні, термодинамічні характеристики та фононна дисперсія орторомбічної фази халькогенідного перовскіту CaZrS$_3$ (просторова група \emph{Pnma}) досліджувалися за допомогою першопринципних  розрахунків із використанням псевдопотенціалів плоскої хвилі в узагальнених градієнтних наближеннях. Були обчислені такі властивості основного стану, як параметри ґратки, об’єм елементарної комірки, модуль всестороннього стиску та його похідна, що добре узгоджуються з результатами, наявними в науковій літературі. Механічні властивості, такі як модулі всестороннього стиску і зсуву, модуль Юнга та пружна анізотропія, були розраховані на основі отриманих констант пружності. Значен\-ня частки модуля всестороннього стиску та модуля зсуву підтверджує, що орторомбічна фаза в CaZrS$_3$ є пластичним матеріалом. Відсутність від'ємних частот на кривій фононної дисперсії та фононна густина станів вказують на те, що ця структура є динамічно стійкою. Насамкінець, були обчислені такі термодинамічні параметри, як вільна енергія, ентропія та теплоємність в залежності від температури. Результати проведених оцінок виявляють такі ж закономірності, як і попередні.
	\keywords CaZrS$_3$, електронні властивості, механічні властивості, дисперсія фононів, термодинамічні властивості
	
\end{abstract}

\lastpage
\end{document}